# Automatic Retrosynthetic Pathway Planning Using Template-free Models

Kangjie Lin[1], Youjun Xu[2], Jianfeng Pei[2,*], Luhua Lai[1, 2, *]

[1]BNLMS, State Key Laboratory for Structural Chemistry of Unstable & Stable Species, College of Chemistry & Molecular Engineering, Peking University, Beijing, 100871, PR China

[2]Center for Quantitative Biology, Academy for Advanced Interdisciplinary Studies, Peking University, Beijing, 100871, PR China

**ABSTRACT**: We present an attention-based Transformer model for automatic retrosynthesis route planning. Our approach starts from reactants prediction of single-step organic reactions for given products, followed by Monte Carlo tree search-based automatic retrosynthetic pathway prediction. Trained on two datasets from the United States patent literature, our models achieved a top-1 prediction accuracy of over 54.6% and 63.0% with more than 95% and 99.6% validity rate of SMILES, respectively, which is the best up to now to our knowledge. We also demonstrate the application potential of our model by successfully performing multi-step retrosynthetic route planning for four case products, i.e., antiseizure drug Rufinamide, a novel allosteric activator, an inhibitor of human acute-myeloid-leukemia cells and a complex intermediate of drug candidate. Further, by using heuristics Monte Carlo tree search, we achieved automatic retrosynthetic pathway searching and successfully reproduced published synthesis pathways. In summary, our model has achieved the state-of-the-art performance on single-step retrosynthetic prediction and provides a novel strategy for automatic retrosynthetic pathway planning.

INTRODUCTION

Organic synthesis, with a history of development over 190 years since the synthesis of urea by Friedrich Wöhler in 1828, is still a rate-limiting step for the discovery of novel medicines and new materials.[1] One of the critical steps for an efficient and environmental-friendly synthesis of valuable molecules lies in the well-designed and feasible retrosynthetic routes. Retrosynthetic analysis, first used by Robert Robinson in tropinone synthesis,[2] then formalized by E. J. Corey,[3] is a fundamental technique that organic chemists applied for designing their target molecules. However, the synthesis route of a molecule is usually diverse, especially for complex compounds like natural products. Planning an efficient and environmental-friendly route of target molecule somehow largely relies on the knowledge of experienced chemists.

The earliest retrosynthesis program could date back to Corey's early work on LHASA (Logic and Heuristics Applied to Synthetic Analysis).[4] Since the 1960s, computer-aided retrosynthetic analysis tools have attracted the interest of many chemists. Computer-aided retrosynthetic design has been well-reviewed over the past years.[5-10] According to a recent review from Coley et al.,[11] the computer-aided retrosynthetic route planning strategies can be clustered into two main categories: template-based and template-free methods. Template-based methods have been applied since the philosophy of retrosynthetic analysis was put forward by E. J. Corey, including the LHASA software developed by Corey and coworkers.[4,12] One of the most known expert-encoded retrosynthetic analysis tools recently developed is Synthia (formerly Chematica), a commercial program developed by Grzybowski and co-workers,[9,13-15] which uses a manually collected knowledge database containing about 70,000 hand-encoded reaction transformation rules. Synthia has been validated experimentally as an efficient toolkit for complex products.[16] Synthia relies on human knowledge of organic synthesis and the encoding of organic rules,[17] which took their team for more than 15 years. It will not be practical to manually collect all the knowledge of organic synthesis considering the exponential growth of the number of published reactions.[18] Another straightforward strategy of template-based method such as the ReactionPredictor from Baldi's group[19-21], is based on mechanistic views. They considered the reactions between reactants as electron sinks and sources and ranked the interactions using approximate molecular orbitals (MOs). Though this kind of approaches were logical and interpretable for chemists, the manual encoding of mechanistic rules cannot be avoided and the mechanisms outside the knowledge database cannot be predicted.

Besides manual rules, automated reaction templates have been extracted by several groups. Based on the algorithms described by Law et al.[22] and Bogevig et al.,[23] Segler and Waller employed a neural network to score templates and perform retrosynthesis and reaction prediction.[24,25] Coupled with Monte Carlo Tree Search (MCTS), they built a novel method for synthetic pathway planning.[18] Later, Coley et al.[26] used the automated extracted templates to perform retrosynthesis analysis based on molecular similarity, where they considered both the similarity of products and reactants to score and rank the templates. However, there are two unavoidable limitations when using automated extracted reaction templates. First, there is an inevitable trade-off between generalization and specificity in template-based methods. Second, it does not consider chemical environment of molecules, as current template extraction algorithm only considers reaction centers and their neighboring atoms. Moreover, mapping the atoms between products and reactants is still an unsolved problem for all template-based methods.[27]

Recently, template-free route planning emerges as a promising strategy. The first template-free model was proposed by Liu and coworkers,[27] using a sequence-to-sequence (seq2seq) model to predict reactants SMILES[28] strings given single products SMILES strings. They used a neural network architecture that involves a bidirectional LSTM encoder and a LSTM decoder with an additive attention mechanism. However, only comparable predicting accuracy was achieved compared to their template-based baseline (65.1% versus 61.7% in top-10 accuracy). Meanwhile, the invalidity rate of the top-10 predicted SMILES strings is over 20%, which restricts the application potential in further synthetic pathway planning.

In 2017, Vaswani et al. proposed an attention-based Transformer model on machine translation tasks[29] and achieved a state-of-the-art performance. Later, two emergent works used this model to predict reaction outcome and reactants for single retrosynthetic analysis.[30,31] Herein, we present a novel template-free strategy using Transformer model to perform automatic retrosynthetic route planning. Trained on a common benchmark dataset (50,000 reactions) from the United States Patent and Trademark Office (USPTO) with known reaction classification information, we achieve a top-1 predictive accuracy of 54.6%, which is superior to the previous template-based or RNN-based seq2seq models.[32] This approach is an end-to-end and data-driven system without considering atom mapping and template extractions. Our model has a powerful capability of generating less invalidity error of SMILES compared to the previous seq2seq model. When applied recursively, our model successfully performed multi-step retrosynthetic route planning in four case examples. More importantly, by using Monte Carlo tree search coupled with a heuristic scoring function, our model can automatically reproduce the above four published pathways, demonstrating the potential of automatic retrosynthetic pathway planning using our novel template-free model.

**METHODS AND MATERIALS**

Cadeddu et al.[33] had described retrosynthesis as natural language processing and termed chemical linguistics. Similarly, retrosynthetic analysis can also be treated as a machine translation problem, where the SMILES strings are considered as sentences and each token or character is treated as a word. In translation, each sentence has several different ways of representations. Similar to that, each product SMILES string can be "translated" to several different reactant SMILES strings, consistent with different disconnections in retrosynthetic analysis. Our seq2seq approach is based on the Transformer architecture, which represents one of the state-of-the-art techniques in neural machine translation.[29] Different from previous (RNN)-based seq2seq model, this architecture is solely based on self-attention mechanisms, which have two main advantages: i) It can significantly improve the efficiency of training time using parallelizable computation, ii) It allows the encoder and decoder to peek at different tokens simultaneously, enabling effective computing for long-range dependent sequences and contributing to producing high-validity SMILES strings.

**Model architecture**. The Transformer architecture[29], depicted in Figure 1, follows an encoder-decoder structure using stacked self-attention and point-wise, fully connected layers. The encoder maps an input symbol sequence $(x_1, \ldots, x_n)$ to a continuous representations $z = (z_1, \ldots, z_n)$. Given $z$, the decoder then generates an output symbol sequence $(y_1, \ldots, y_m)$. The encoder and decoder are composed of a stack of $N$ identical layers, each of which contains three sub-modules. The first sub-module is a multi-head self-attention mechanism, which is made of several scaled dot-product attention layers running in parallel. In this attention layer, the input consists of queries $Q$ and keys $K$ of $d_k$ dimension, and values $V$ of $d_v$ dimension. The formula of a single attention function is

$$\text{Attention}(Q, K, V) = \text{softmax}(\frac{QK^T}{\sqrt{d_k}})V$$

This attention function will yield $d_v$-dimensional output values. These values are concatenated into the multi-head attention layer with the formula

$$\text{MultiHead}(Q, K, V) = \text{Concat}(\text{head}_1, \ldots, \text{head}_h)W^O$$
$$\text{where } \text{head}_i = \text{Attention}(QW_i^Q, KW_i^K, VW_i^V)$$

Where, the trainable parameter matrices, $W_i^Q \in \mathbb{R}^{d_{\text{model}} \times d_k}, W_i^K \in \mathbb{R}^{d_{\text{model}} \times d_k}, W_i^V \in \mathbb{R}^{d_{\text{model}} \times d_v}, W^O \in \mathbb{R}^{d_{\text{model}} \times hd_k}$, are the linear projections. The number of parallel attention layers or heads is $h$. For each of these, $d_k = d_v = d_{\text{model}}/h$.

The second sub-module is a fully connected feed-forward network, which is applied to each position separately and identically. The transformation function is

$$FFN(x) = max(0, xW_1 + b_1)W_2 + b_2$$

Where $W_1, W_2, b_1, b_2$ are learnable weights and biases.

The third sub-module in the decoder stack is a modified self-attention layer using masking operation to prevent positions from attending to subsequent positions, which ensures that predictions at position $i$ can be only up to the known outputs at position $< i$.

Similar to other sequence models, the embedding layer is added to convert the input tokens and output tokens to $d_{\text{model}}$-dimension vectors. To preserve the order of the sequence, positional encoding operation is combined into the input embeddings, and these encodings have the same dimension $d_{\text{model}}$ as the embeddings, so that the two can be summed. The sine and cosine functions of different frequencies are used in the positional encoding as follow

$$PE_{(pos,2i)} = sin(pos/10000^{2i/d_{\text{model}}})$$

$$PE_{(pos,2i+1)} = cos(pos/10000^{2i/d_{\text{model}}})$$

Where $pos$ is the position and $i$ is the dimension.

**Datasets**. In our work, two datasets were used to develop our single-step retrosynthetic prediction models. First, a common benchmark dataset with 50,000 reactions (called USPTO_50K) was previously applied by Liu et al.[27] and Coley et al.[26] The reaction classes in the dataset have been labelled by Schneider and coworkers as described in Table 1. Following Liu et al., we used a 90%/10% training/testing split and the validation set was randomly sampled from training sets (10%). Second, in order to develop a more powerful model, we also used a much larger dataset called USPTO_MIT[34] from the USPTO[32] with a pre-processed training, validation, and testing sets of 424573, 42457 (randomly sample from training sets), 38648 reactions, respectively.

**Data preprocessing**. Inspired by Schneider et al.[35,36], the original USPTO_MIT dataset was preprocessed to extract the reactants and products of each reaction. We classified the reactions using reaction fingerprints and agent features. The details of the reaction classification algorithm and results can be referred to Supporting Information. Moreover, we tried token- and character-based methods to tokenize the SMILES strings as model input. The difference between token- and character-based preprocessing is described in Supporting Information.

**Monte Carlo Tree Search**. To implement automatic search, MCTS[37] is used to create a search tree, where each node corresponds to a set of molecules (shown in Figure 2). Nodes with the terminal molecules (precursor molecules) are called terminal nodes. Starting with the root node (a target molecule), the search tree grows gradually by iterating the four steps, including selection, expansion, simulation and backpropagation. Each intermediate node has an upper confidence bound (UCB) score that is an index indicating how promising it is to explore the subtree. For the rollout in the simulation step, when a new node is added, paths from the expanded node to terminal nodes are built by a random approach. Instead of a uniformly random rollout, the well-trained retrosynthesis predictive models were employed to design the rollout procedure for a better and faster searching. A node at $t-1$ has a partial retrosynthesis pathway $(s_1, \ldots, s_{t-1})$ corresponding to the path from the root to this node. Based the node $s_{t-1}$, our model allows us to compute the distribution of the next node $s_t$. Sampling from the distribution, the pathway is elongated by one step. Elongation by our model is repeated until the terminal node occurs. Once elongation is done, the defined reward score of the generated pathway is used to propagate backward for updating the UCB scores of traversed nodes during the backpropagation process.

**Code implementation**. All program scripts were written in Python (version 3.6), and open source RDKit (version 2018.09.02)[38] was used for reaction preprocessing and SMILES validation. Our seq2seq model was built with TensorFlow (version 1.12.0)[39] and the details of key hyperparameters settings of our models are available in Supporting Information.

## RESULTS

**Single-Step Evaluation**. As summarized in Table 2, we have achieved best top-1 accuracy of 54.6% and 63.0% in USPTO_50K and USPTO_MIT dataset, respectively. With prior reaction classes information, the top-1 prediction accuracy of our model is much better than the LSTM-based seq2seq model proposed by Liu et al.[27] and also higher than the template-based model by Coley et al.[26] When the reaction classes are not provided, our model still significantly outperforms the template-based model of Coley et al. by 5.8%.

The comparison of the top-10 accuracy across all classes of our model with the previous works on USPTO_50K and USPTO_MIT dataset is shown in Table 3. The performance of our model is much better than the seq2seq model of Liu et al. across all reaction categories. However, our model just performs slightly better or comparable to the template-based model in category 3 (C-C bond formation), 7 (reductions) and 9 (functional group interconversion). This might be due to the imbalanced data of each category.

The ratio of invalid SMILES strings produced by our model is much lower than the previous LSTM-based model, which means that our model has a powerful ability of capturing the grammar of SMILES representations. As shown in Table 4, the top-10 invalidity error of our model is 12.7%, even close to the top-1 invalidity error of Liu's model. When we trained our model on the large-volume USPTO_MIT dataset, the top-1 accuracy increases to 63.0%, which shows the generality ability of our model by increasing chemical knowledge base. Similarly, worse performance also exists due to imbalanced data of each class. Meanwhile, the error rate of SMILES strings decreases to 8.5% in top-10 prediction.

As shown in Figure 3, we used the top 5 retrosynthetic disconnections of a compound in test set as an example to analyze the specificity and generality of our model. We chose a compound in class 1 as an example, in which the ground truth prediction ranks first, and the other predicted reactions are also chemically plausible. It shows that our model is able to give reasonable diverse disconnections. Remarkably, the top 5 disconnections comply with the reaction class of heteroatom alkylation.

**Iterative Multi-Step Pathway Generation.** As the prediction accuracy of our model is quite high (even higher than template-based methods), we also tried the potential of our model in recursive generation of candidate reactants. We took four target compounds as examples, including antiseizure drug Rufinamide,[40] a novel allosteric activator for GPX4, and other two representative compounds used by other retrosynthetic programs[16,18]. By enumerating different reaction classes, our model could successfully reproduce the published reaction pathways of the four compounds.

For the first example of retrosynthesis pathway planning of Rufinamide (shown in Figure 4), the reported first step is the formation of amide bond, ranking first in reaction class 9 (functional group interconversion). The subsequent step is also found to rank top-1 in class 4 (heterocycle formation), consistent with the mechanistic view. Followed by another functional group interconversion (FGI) step, the final step is predicted precisely as class 9 in top-1. It is worth to be mentioned that different reaction class may have the same disconnections and thus resulting in the same reactants. For example, the third step of the above route also ranks first in class 1 (heteroatom alkylation), which is also plausible in chemistry.

For the second example of retrosynthesis pathway planning of the GPX4 activator compound, as depicted in Figure 5. The published first step ranks second in class 5 (acylation and related process). The second step could be regarded as functional group interconversion (FGI) and it is predicted correctly as top-1 by our model. The ground truth of third step ranks top-2 in class 10, preceded by a final alkylation step in top1 class 1.

As shown in Figure 6, the third example comes from the previous work of Grzybowski et al.,[16] which is the retrosynthesis pathway planning of an antagonist of the interaction between WD repeat-containing protein 5 (WDR5) and mixed-lineage leukemia 1 (MLL1)[41], our model could recover that route suggested by commercial program Synthia. The first step is a function group interconversion (FGI) predicted as top-1 by our model, followed by a common amide formation. The final step is a C-C bond formation and also predicted by our model as top-1 with correct reaction class.

The fourth example, described in Figure 7, is the retrosynthesis pathway planning of an intermediate of drug candidate from the example of Segler and coworkers.[12] The reported route cannot be completely predicted by our model trained on USPTO_50K dataset due to less coverage of chemical space. Remarkably, when trained on USPTO_MIT dataset, our model could completely reproduce the six-step route in our top-10 predictions, suggesting the importance of training on enlarging coverage of chemical knowledge space. The first third step can be easily reproduced by our model as top-1 with right class. The fourth step is a common functional group addition (FGA), followed by an uncommon reductions of a carbonyl group. The fifth step is also the toughest step in the total steps of all four examples. After the final step of heteroatom alkylation, our model could reproduce the steps predicted by the former template-based method.

**Automatic Retrosynthetic Pathway Planning.** As shown above, when considering the top 10 prediction of each of the 10 reaction classes, 100 candidate reactants for a target will be predicted in one step. A recursive application in a 4-step pathway will produce 100,000,000 candidate pathways assuming all of the output SMILES strings are valid. To make our model applicable for retrosynthetic pathway planning, we need to achieve efficient automatic pathway searching and ranking. We used a Monte Carlo tree search algorithm combined with a heuristic scoring function to achieve this purpose. Our heuristic scoring function is inspired from Synthia's Chemical Scoring Function (CSF).[9] We take the $Score_{model}$, representing the decoding log probability from the beam search, produced by our model, the changed SMILES length from target to the reactants and the changed number of rings from target to reactants into consideration. To scale the heuristic scoring function in a comparable range, we presented the function in a formula of $a \times \exp(Score_{model}) - (b \times RINGS_{changed} + SMILES_{changed})$. We define the parameters a, b as 100, 6 in our four examples.

We used a dataset containing 84,807 building blocks from chemical suppliers (SigmaAldrich), obtained from the ZINC15 database (http://zinc15.docking.org/) and 17,182 molecules from the USPTO_MIT database, used as reactants at least five times as terminal nodes (a building blocks database of 93,563 molecules after removing redundant ones) for searching. Users can also use any specific building block database as terminal reactant database.

Using our automatic retrosynthetic pathway planning strategy, most of the aforementioned steps in the four case examples can be found and ranked in top-10 except for the first step of example 2 (ranks top-11) and fifth step of example 4 (ranks top-25). The overall pathway ranking results of four case examples can be found in Supporting Information. Though our heuristic scoring function is simple, these results are impressive, demonstrating the potential ability of our template-free model to plan automatic retrosynthetic pathway, a new way other than by using current template-based methods.

**DISCUSSION**

    **Advantages and disadvantages of our seq2seq models.** As shown above, our models are template-free and free of atom mapping. Besides, our models can learn the global chemical environments of molecules naturally compared to other template-based methods. However, they still have some problems related to dataset and SMILES representations. Apart from less coverage of chemical reaction space, the USPTO dataset does not contain reaction yield information for reactions, which is useful to discriminate whether the predicted pathways are efficient. Because our models were trained on USPTO datasets, their prediction accuracies are currently limited by these problems. A commonly known challenge of using SMILES or reaction SMARTS format is the poor performance when dealing with stereochemistry and tautomers. Like other template-based methods, our models are still difficult to tackle the reactions containing chirality. In fact, our models are able to treat reactants or products with simple chirality as long as we include the reactions containing chirality. However, as mentioned by Grzybowski et al.[9] and Segler et al.[12], this problem is related to the grammar of SMILES/SMARTS due to the lack of keeping track of the changes in chirality. Furthermore, since our models do not contain any information about reaction conditions, it is currently impossible to deal with asymmetric synthesis, most of which rely on asymmetric catalysts. Meanwhile, tautomers, though chemically equivalent in different molecular structures, are regarded as different inputs and outputs in our model because current SMILES grammars are sequence sensitive. This is also a common problem in template-based model as described by Segler et al.[12] Embedding stereochemistry and tautomerization into SMILES representation is a future direction.

    **Evaluation of different pathways.** There are different pathways predicted by retrosynthetic programs. One model can offer thousands of different pathways. However, picking a suitable pathway from them is not easy. For medicinal chemists, they may want a pathway expanding structure activities relationship exploration. For organic chemists especially those working in total synthesis of natural products, they may have preference to the more efficient and greener pathways. For process chemists, the cost of starting materials and avoidance of toxic and active molecules may influence their final choices. It is difficult to find a pathway that fulfill all these requirements. A heuristic metric proposed by Synthia seems to be a reasonable strategy.[9] They considered two scoring functions: Chemical Scoring Function (CSF) and Reaction Scoring Function (RSF). Another potential strategy is to use the SCScore metric proposed by Coley and coworkers.[42] In general, a comprehensive scoring function is related to the cost of building blocks, the yield of each step, the avoidance of toxic compounds and functional group incompatibility, the length of the pathway, etc. The design of a perfect pathway scoring function is still an unsolved problem in the community.

    **Evaluation of different models.** Evaluation of retrosynthetic analysis approach is difficult by using a benchmark metrics. The strategy applied by Segler et al.[12] is a reasonable one. They invited professional organic chemists to vote the predicted and ground truth pathways. If the chemists do not show preference to the ground truth pathway, it means that the predicted one is also reasonable. However, this assessment is difficult to be standardized. Certainly, validation in wet lab is the most reliable way. As most chemists are interested in the synthesis of novel complex compounds or finding efficient alterative pathways for valuable molecules, validation of these kinds of compounds with wet

experiments should be considered. For example, the cooperation between MilliporeSigma and Grzybowski et al. had resulted in the efficient syntheses of eight diverse and medicinally relevant targets, making the chemistry community realize the reliability of Synthia.[9]

## CONCLUSION

In this work, we have demonstrated a novel approach for retrosynthetic analysis using Transformer-based seq2seq model. By enumerating ten reaction classes and predicting top 10 disconnections of each class, our model could reproduce four published retrosynthetic pathways. To further demonstrate the potential of our model to perform automatic retrosynthetic route planning, we applied Monte Carlo Tree Search combined with a heuristic scoring function to explore the potential routes for a given target molecule. The published pathways of above four examples could be recovered using our heuristic MCTS method. Unlike other template-based methods, which either relied on experts' laborious work or simple contextless rule-based system, our approach is fully end-to-end and incorporates the global molecular context of the reaction species naturally. We have first shown that a template-free approach can be used to perform automated retrosynthetic pathway searching and could reproduce the published synthesis pathway of valuable compounds. As mentioned by Coley et al., a complete retrosynthetic program is made up of five components[11]: a library containing the disconnections rules; a recursive application engine that generates candidate reactants for target compounds; a building blocks database containing available compounds acts as terminal nodes; a strategy to guide the retrosynthetic search; a scoring function for single-step or pathway. And our approach have included all of these components as shown in above.

Our approach may play an important role in retrosynthetic route planning with larger and more diverse chemical knowledge base. After all, the information of reaction conditions like catalysts, solvents and reagents are missing in our model due to the lack of a more comprehensive datasets like Reaxys database or in-house electronic lab notebook data. Future work is also required to tackle the problems like SMILES' poor representations in stereochemistry and tautomerization. Furthermore, we believe that potential application of retrosynthetic program may play an important role in de novo molecular design[43] and automated synthesis of molecules[44].

## ACKNOWLEDGMENT

This work has been supported by the National Science & Technology Major Project "Key New Drug Creation and Manufacturing Program", China (2018ZX09711002); the National Natural Science Foundation of China (21673010, 21633001); and the Ministry of Science and Technology of China (2016YFA0502303). Part of the analysis was performed on the High Performance Computing Platform of the Center for Life Science.

## AUTHOR INFORMATION

**Corresponding Author**

*Fax: (+86)10-62759595. E-mail: jfpei@pku.edu.cn.

*Fax: (+86)10-62751725. E-mail: lhlai@pku.edu.cn.

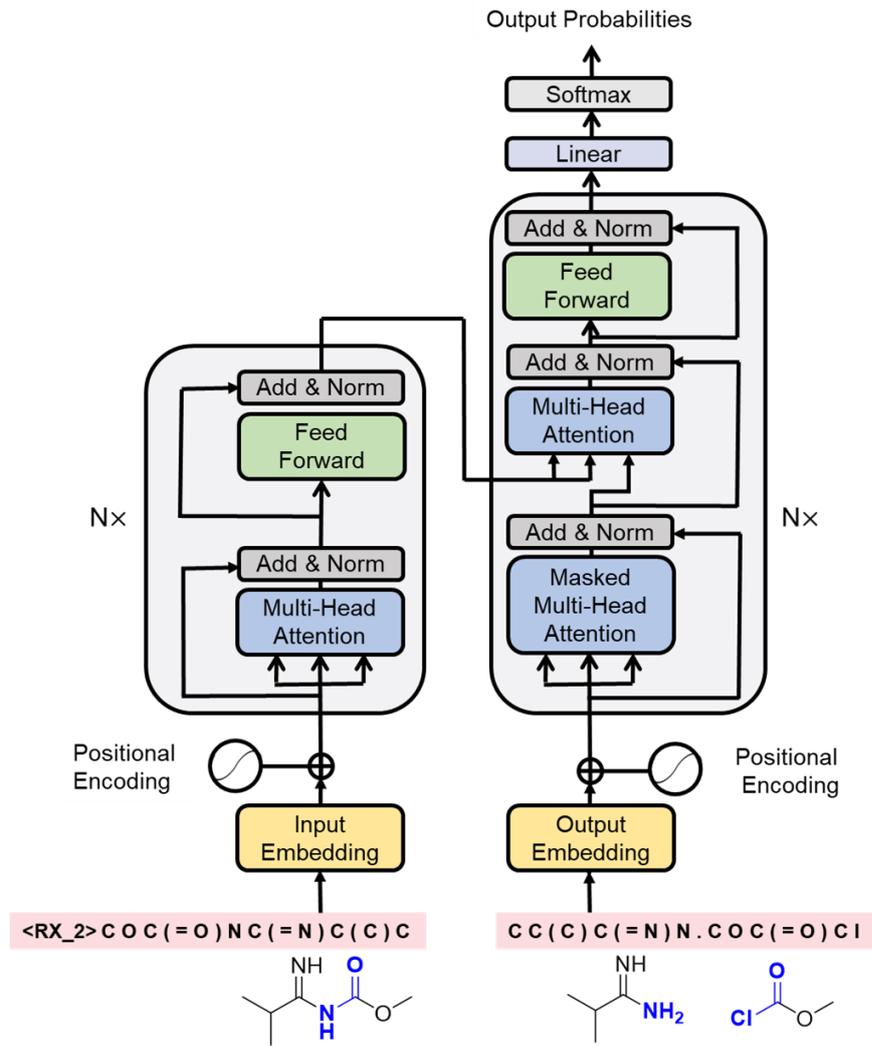

Figure 1. Transformer model architecture and explanation of training process.

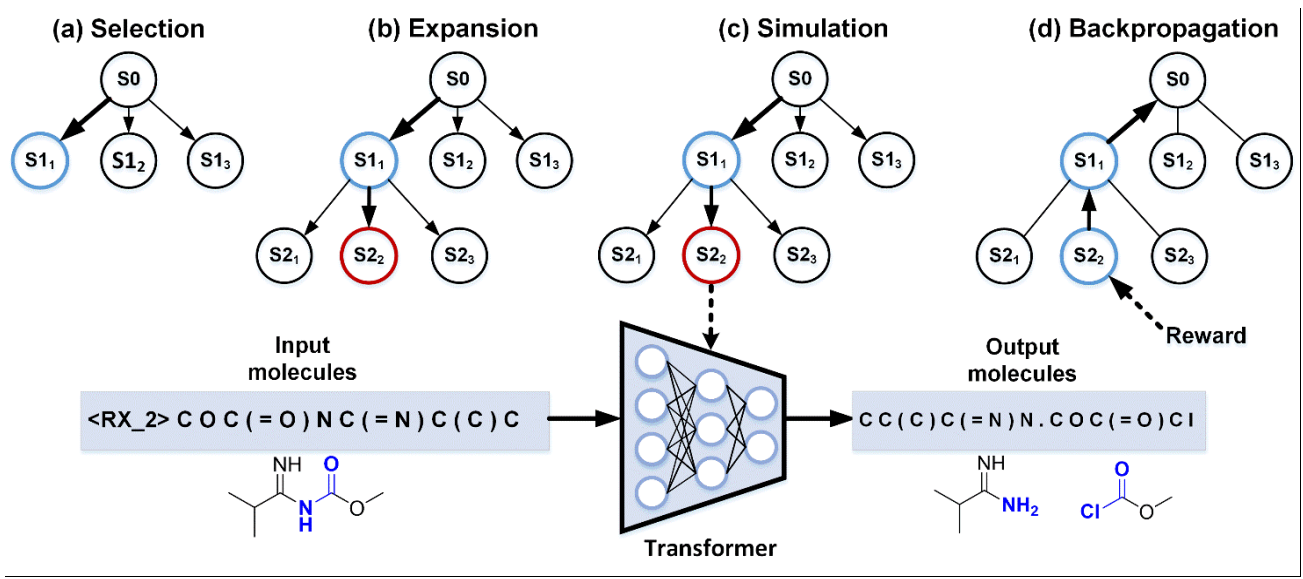

Figure 2. Monte Carlo Tree Search and its application in retrosynthetic analysis. (a) Selection step. The search tree is traversed from the root to a leaf by choosing the child with the largest UCB score. (b) Expansion step. (c) Simulation step. Paths to terminal nodes are created by the rollout procedure using Transformer model. Rewards of the corresponding molecules are computed. (d) Backpropagation step. The internal parameters of upstream nodes are updated.

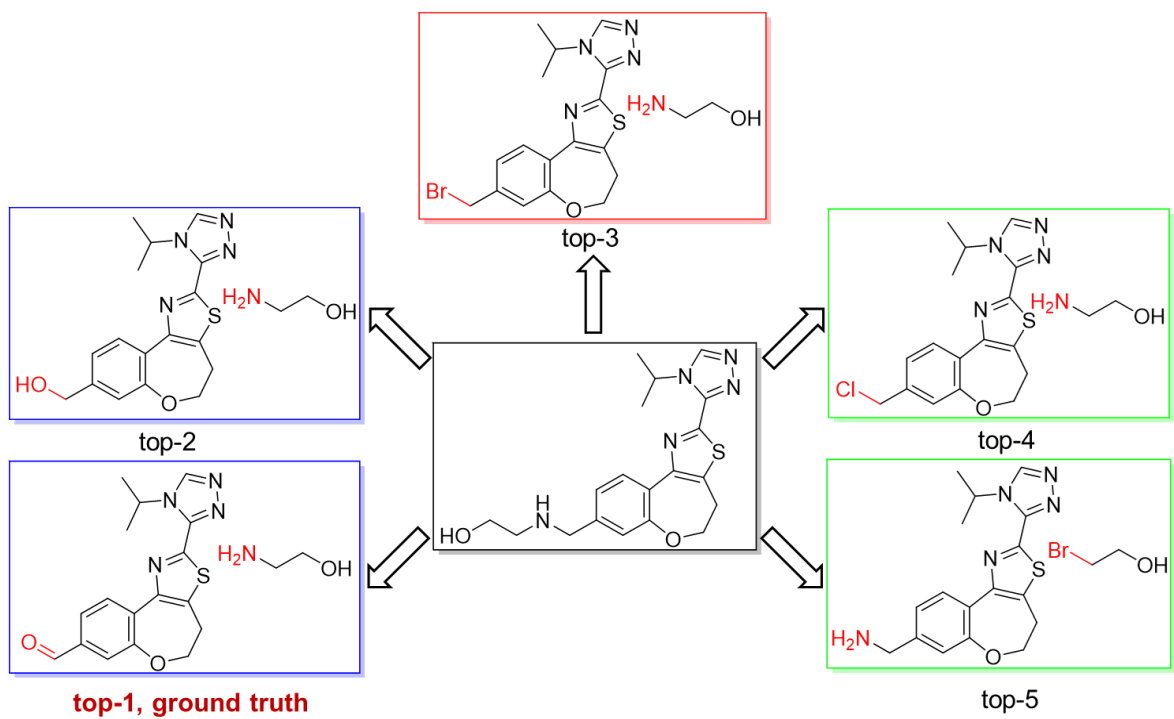

Figure 3. top-5 retrosynthetic predictions of an example reaction in class 1. The model successfully proposes the recorded reactants with rank 1, corresponding to a heteroatom alkylation. Other suggestions among the top 5 predictions are also grammatically valid while chemically plausible.

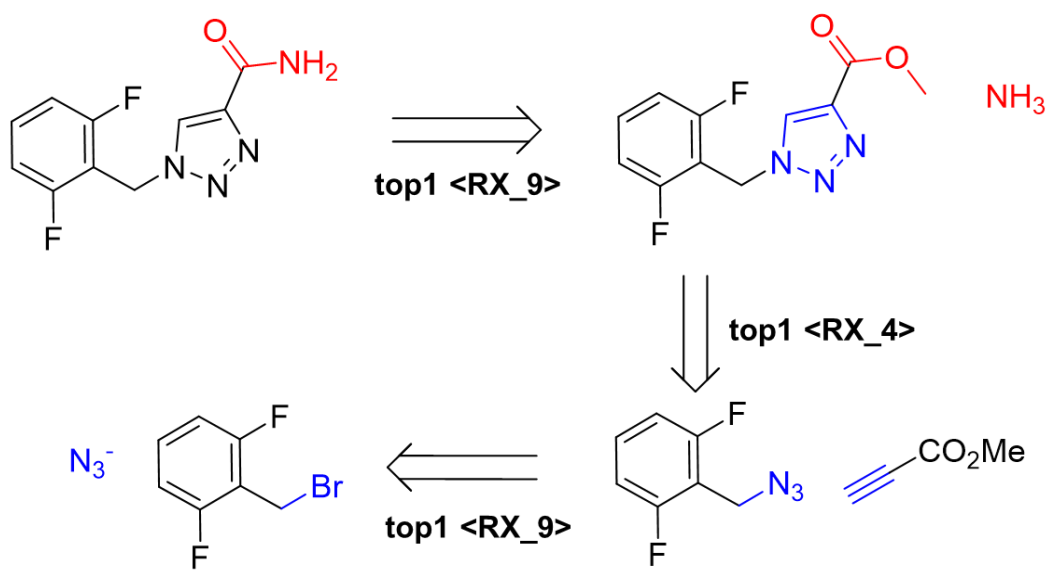

Figure 4. Iterative multi-step pathway generation. Routes are constructed by iteratively applying the single-step retrosynthetic methodology to Rufinamide. The suggested disconnections are consistent with published pathways.

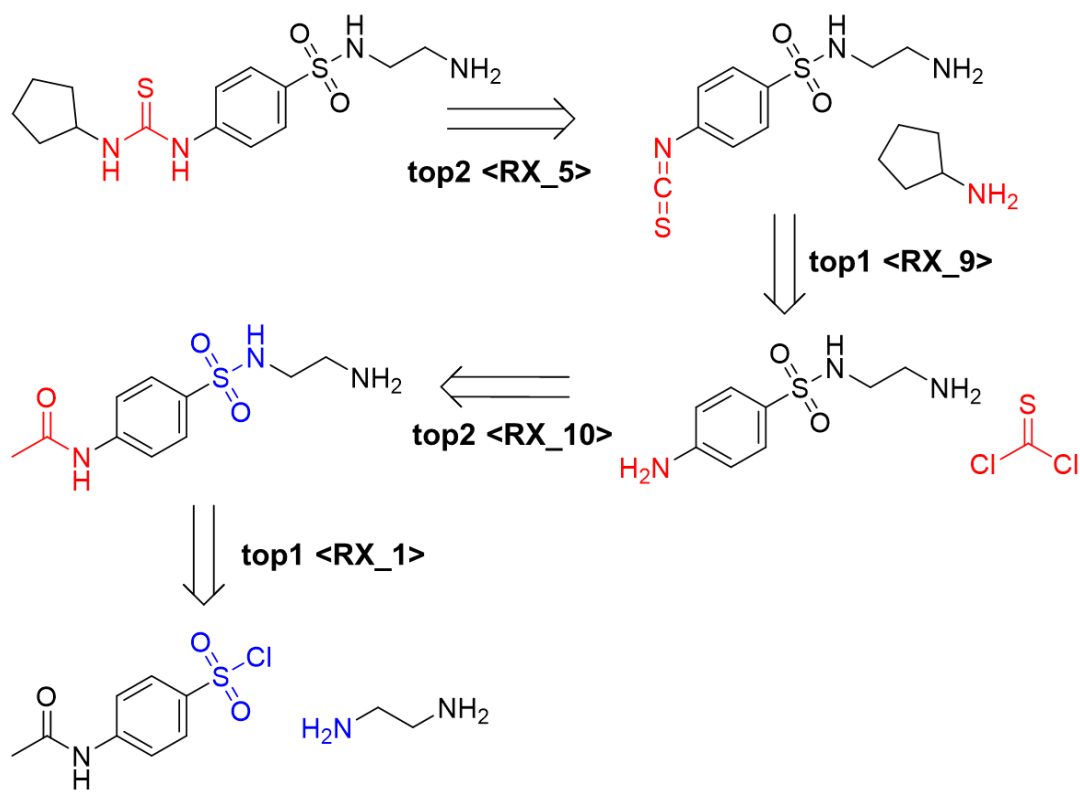

Figure 5. Iterative multi-step pathway generation. Routes are constructed by iteratively applying the single-step retrosynthetic methodology to an allosteric activator for GPX4. The suggested disconnections are consistent with published pathways.

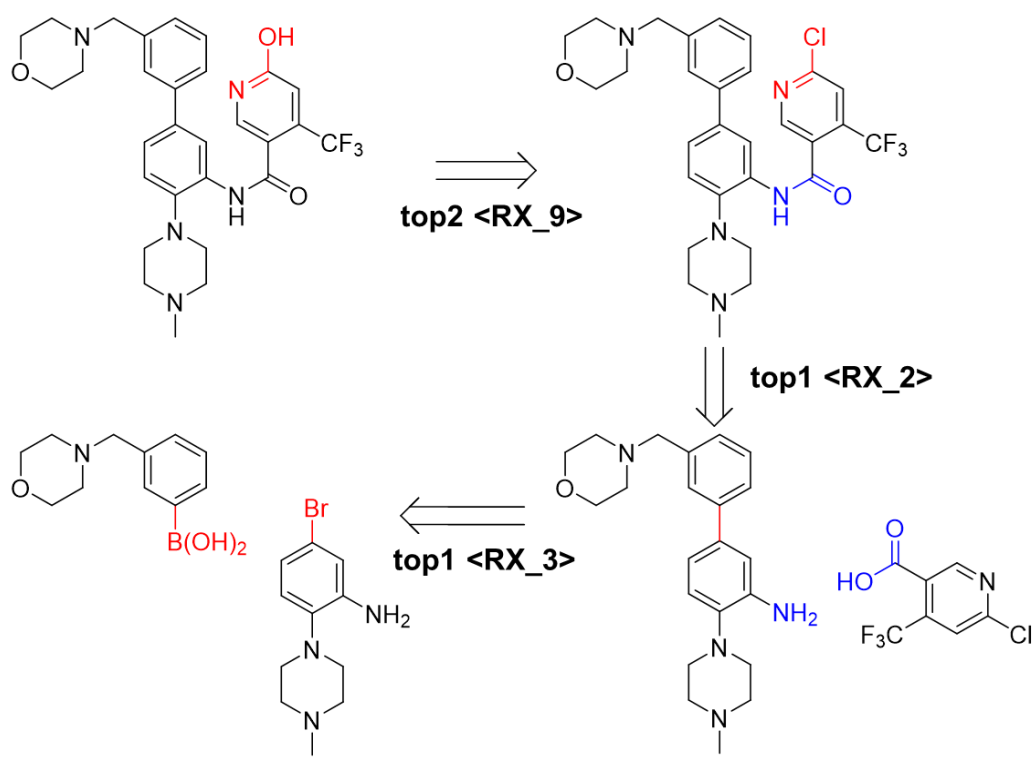

Figure 6. Iterative multi-step pathway generation. Routes are constructed by iteratively applying the single-step retrosynthetic methodology to an antagonist of the interaction between WDR5 and MLL1, from the examples of Grzybowski et al. The suggested disconnections are consistent with published pathways.

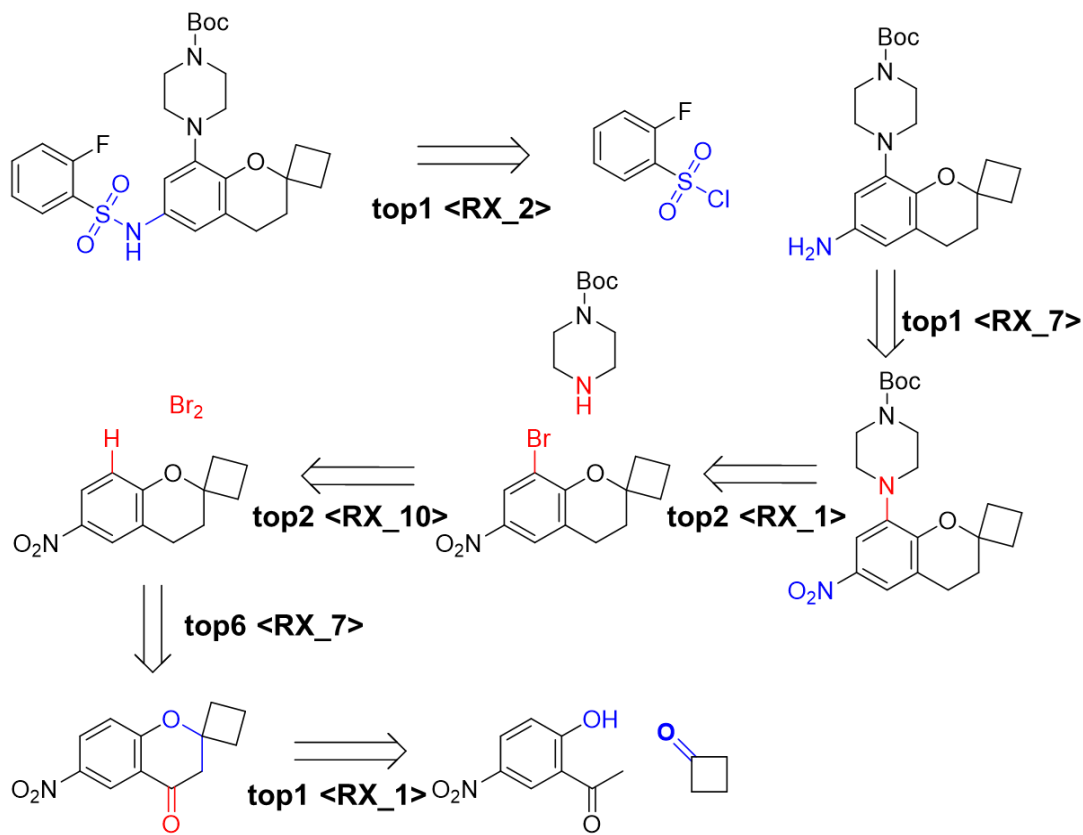

Figure 7. Iterative multi-step pathway generation. Routes are constructed by iteratively applying the single-step retrosynthetic methodology to an intermediate of drug candidate from the examples of Segler et al. The suggested disconnections are consistent with published pathways.

Table 1. Descriptions of Each of the 10 Reaction Classes and the Fraction of the USPTO_50K and USPTO_MIT

| reaction class | reaction name | fraction of USPTO_50K (%) | fraction of USPTO_MIT (%) |
| --- | --- | --- | --- |
| 1 | heteroatom alkylation and arylation | 30.3 | 29.9 |
| 2 | acylation and related processes | 23.8 | 24.9 |
| 3 | C-C bond formation | 11.3 | 13.4 |
| 4 | heterocycle formation | 1.8 | 0.7 |

| 5 | protections | 1.3 | 0.3 |
| 6 | deprotections | 16.5 | 14.1 |
| 7 | reductions | 9.2 | 9.4 |
| 8 | oxidations | 1.6 | 2.0 |
| 9 | functional group interconversion (FGI) | 3.7 | 5.0 |
| 10 | functional group addition (FGA) | 0.5 | 0.2 |

**Table 2. Model Performance Aggregated Across All Classes**

| | top-$n$ accuracy (%), $n =$ | | | |
|---|---|---|---|---|
| model | 1 | 3 | 5 | 10 |
| Liu et al. baseline (USPTO_50K) | 35.4 | 52.3 | 59.1 | 65.1 |
| Liu et al. seq2seq (USPTO_50K) | 37.4 | 52.4 | 57.0 | 61.7 |
| Similarity+class (USPTO_50K) | 52.9 | 73.8 | **81.2** | **88.1** |
| Similarity (USPTO_50K) | 37.3 | 54.7 | 63.3 | 74.1 |
| Transformer+token (USPTO_50K) | 42.0 | 64.0 | 71.3 | 77.6 |
| Transformer+token+class (USPTO_50K) | 54.3 | 74.1 | 79.2 | 84.4 |
| Transformer+char (USPTO_50K) | 43.1 | 64.6 | 71.8 | 78.7 |
| Transformer+char+class (USPTO_50K) | **54.6** | **74.8** | 80.2 | 84.9 |
| Transformer+char (USPTO_MIT) | 54.1 | 71.8 | 76.9 | 81.8 |
| Transformer+char+class (USPTO_MIT) | 63.0 | 79.2 | 83.4 | 86.8 |

**Table 3. Model Top-10 Accuracy within Each Class When the Reaction Class Is Known a Priori**

| | reaction class, top-10 accuracy (%) | | | | | | | | | |
|---|---|---|---|---|---|---|---|---|---|---|
| model | 1 | 2 | 3 | 4 | 5 | 6 | 7 | 8 | 9 | 10 |
| Liu et al. baseline (USPTO_50K) | 77.2 | 84.9 | 53.4 | 54.4 | 6.2 | 26.9 | 74.7 | 68.4 | 46.7 | 73.9 |
| Liu et al. seq2seq (USPTO_50K) | 57.5 | 74.6 | 46.1 | 27.8 | 80.0 | 62.8 | 67.8 | 69.1 | 47.3 | 56.5 |

| Similarity+class (USPTO_50K) | **86.7** | **94.2** | 74.6 | **67.0** | **97.1** | **95.5** | **88.3** | **98.8** | 71.2 | **91.3** |
| Transformer+char+class (USPTO_50K) | 83.1 | 90.4 | **76.2** | 60.0 | 92.3 | 88.6 | 88.2 | 86.4 | **73.9** | 82.6 |
| Transformer+char+class (USPTO_MIT) | 88.2 | 91.2 | 81.9 | 67.8 | 75.4 | 86.6 | 87.1 | 88.5 | 73.5 | 66.7 |

**Table 4. Breakdown of the Grammatically Invalid SMILES Error for Different Beam Sizes**

| | invalid SMILES rate (%) | | | |
| --- | --- | --- | --- | --- |
| model | 1 | 3 | 5 | 10 |
| Liu et al. seq2seq (USPTO_50K) | 12.2 | 15.3 | 18.4 | 22.0 |
| Transformer+token (USPTO_50K) | 2.2 | 3.7 | 4.8 | **7.8** |
| Transformer+token+class (USPTO_50K) | 2.3 | 4.9 | 7.0 | 12.1 |
| Transformer+char (USPTO_50K) | **2.1** | **3.5** | **4.7** | 8.3 |
| Transformer+char+class (USPTO_50K) | 2.4 | 4.5 | 6.4 | 12.7 |
| Transformer+char+class (USPTO_MIT) | 0.4 | 1.4 | 2.9 | 8.5 |